\documentclass[twocolumn,trackchanges]{aastex701}
\usepackage{amsmath}
\usepackage{subcaption}
\usepackage{graphicx}

\begin{document}

\title{Simulation-based inference for neutron star pulse profile modeling}

\author[0009-0005-8019-0426]{Mariska~Hoogkamer}
\affil{Anton Pannekoek Institute for Astronomy, University of Amsterdam, Science Park 904, 1098XH Amsterdam, the Netherlands}
\email{m.m.hoogkamer@uva.nl}

\author[0000-0002-2651-5286]{Devarshi~Choudhury}
\affil{Anton Pannekoek Institute for Astronomy, University of Amsterdam, Science Park 904, 1098XH Amsterdam, the Netherlands}
\email{d.choudhury@uva.nl}

\author[0000-0002-1169-7486]{Daniela~Huppenkothen}
\affil{Anton Pannekoek Institute for Astronomy, University of Amsterdam, Science Park 904, 1098XH Amsterdam, the Netherlands}
\email{d.huppenkothen@uva.nl}

\author[0000-0002-1009-2354]{Anna~L.~Watts}
\affil{Anton Pannekoek Institute for Astronomy, University of Amsterdam, Science Park 904, 1098XH Amsterdam, the Netherlands}
\affil{Gravitation and Astroparticle Physics Amsterdam (GRAPPA), University of Amsterdam, 1098XH Amsterdam, the Netherlands}
\email{A.L.Watts@uva.nl}

\author[orcid=0000-0002-6449-106X]{Sebastien Guillot}
\affiliation{University of Toulouse, CNES, CNRS, IRAP, Toulouse, France}
\email{sebastien.guillot@utoulouse.fr}



\begin{abstract}
NICER has enabled mass--radius inferences for pulsars using pulse profile modeling (PPM), providing constraints on the equation of state (EOS) of cold, dense matter. A common approach in PPM uses nested sampling to obtain posterior distributions. While well established, this approach is computationally expensive, restricting how many tests and models can be run and how complex those models can be. The cost will become a serious bottleneck for the higher quality data sets expected from future telescopes such as eXTP and NewAthena. 
Here we present an alternative approach based on simulation-based inference (SBI), using truncated sequential neural posterior estimation (TSNPE), and compare it against the nested sampler MultiNest. We apply it to synthetic NICER data generated for two relatively simple neutron star models, in which the surface hot emitting regions are described by either one or two independent circular, uniform-temperature spots. In both cases TSNPE recovers the injected parameter values and yields posteriors consistent with those from MultiNest. Retaining access to the likelihood, we further estimate the Bayesian evidence by neural importance sampling and find agreement with MultiNest to within $\sim0.2-0.4$ in $\ln\mathcal{Z}$ for both models. TSNPE achieves this at a substantially lower computational cost, giving a speed-up factor between $32.7$ and $241.8$ compared to MultiNest, depending on the model. Our results lay the groundwork for applying SBI to real data, and the speed-up opens the door to more complex models and to systematic stress testing.
\end{abstract}


\section{Introduction}\label{sec:introduction}
Neutron stars provide a unique opportunity for studying dense nuclear matter and constraining its equation of state (EOS) \citep[see, e.g.,][]{Lattimer2016-EOS, Baym2018-EOS, Tolos2020, Chatziioannou2025-NS_EOS}. Rotation-powered millisecond pulsars exhibit rotationally modulated, pulsed X-ray emission, believed to originate from surface hot spots at the magnetic poles where return currents deposit heat \citep[see, e.g.,][]{Ruderman1975-pulsars, Arons1981, Harding2001}. This X-ray signal can be analyzed using the pulse profile modeling (PPM) technique \citep[e.g.,][and references therein]{Bogdanov2019-part2, Bogdanov2021-part3}, which incorporates relativistic effects arising from the neutron star’s rapid spin and intense gravitational field, enabling precise measurements of neutron star masses and radii. The Neutron Star Interior Composition Explorer \citep[NICER;][]{Gendreau2016-NICER} has been central to this effort, detecting the soft thermal X-rays from several rotation-powered millisecond pulsars needed for PPM. 

To date, NICER has enabled mass and radius inferences for seven rotation-powered millisecond pulsars: PSR~J0030+0451 \citep{Riley2019-J0030-xpsi, Miller2019-J0030, Salmi2023-atmosphere, Vinciguerra2024-J0030, Kini2026-J0030}, PSR~J0437--4715 \citep{Choudhury2024-J0437, Qi2026-J0437, Miller2026-J0437, Stammler2026-J0437}, PSR~J0740+6620 \citep{Miller2021-J0740, Riley2021-J0740, Salmi2022-J0740, Salmi2023-atmosphere, Salmi2024-J0740, Dittmann2024-J0740, Hoogkamer2025-J0740}, PSR~J1231--1411 \citep{Salmi2024-J1231, Qi2025-J1231}, PSR~J0614--3329 \citep{Mauviard2025, Miller2026-J0614}, PSR~J2124--3358 \citep{Gonzalez-Caniulef2026-J2124}, and PSR~J1614--2230 \citep{Mauviard2026-J1614}. These mass--radius constraints in turn have been widely used to constrain the EOS of cold, dense matter \citep[see, e.g.,][]{Koehn24, Kurkela24, Huang25, Golomb25, Ng25, LiJJ24, Gartlein26, Mendes2026-EOS_J1614}. 

Many of the aforementioned neutron star analyses rely on the open-source software package X-PSI\footnote{\url{https://github.com/xpsi-group/xpsi}} \citep[X-ray Pulse Simulation and Inference;][]{xpsi}, for PPM and Bayesian statistical inference, combined with a nested sampler such as MultiNest \citep{Feroz2008-MultimodalNestedSampling, Feroz2009-Multinest, Feroz2019-Multinest} and its Python binding PyMultiNest \citep{Buchner2016-PyMultiNest}, or UltraNest \citep{Buchner2021-UltraNest}. Although PPM with nested sampling is well-established, it is computationally very expensive. For instance, inference for PSR~J0030+0451 required $\sim4.8$ million core hours to convergence in a single run \citep{Kini2026-J0030}. This cost is a bottleneck, limiting both the number of tests and models that can be run and how complex those models can be. Models incorporating more, or more realistic, physics are especially affected, for example those with physical surface patterns from magnetospheric modeling \citep{Gralla17, Lockhart2019, Huang2026}, or the additional disc and atmosphere components required for accreting millisecond pulsars \citep{Dorsman2025-AMPs, Dorsman2026-SAXJ1808, Dorsman2026-SGRAJ1444, Mao2026-AMPs}. The problem will only grow with the higher quality data sets expected from future telescopes such as eXTP \citep{LiA25} and NewAthena \citep{Cruise25}. As the data become more informative, more complex models will likely be needed to accurately represent the data, which increases the sampling cost. At the same time, the wider energy range enabled by the higher sensitivity, and the finer energy resolution (e.g., with NewAthena's X-ray Integral Field Unit) require more ray tracing. All of these factors raise the computational cost. 

Here, we explore a simulation-based inference (SBI) technique called truncated sequential neural posterior estimation \citep[TSNPE;][]{Deistler2022-TSNPE} to reduce the computational cost of inferring neutron star model parameters relative to nested sampling. TSNPE is a variant of neural posterior estimation \citep[NPE;][]{Papamakarios2016-NPE, Greenberg2019-APT}, in which a neural density estimator is trained on simulated data to approximate the posterior distribution of the underlying parameters. 

NPE is well established in astronomy as a tool for fast, amortized parameter inference. It has been applied, for example, to gravitational wave signals \citep[e.g.,][]{Dax2021-GW_NPE}, binary microlensing events \citep[e.g.,][]{Zhang2021}, dynamical masses of galaxy clusters \citep{KodiRamanah2020}, galaxy spectral energy distributions \citep[e.g.,][]{Hahn2022}, and exoplanet atmospheric retrieval \citep[e.g.,][]{Vasist2023-NPE_exoplanet}. In the neutron star and X-ray astrophysics context specifically, NPE has been developed for the analysis of X-ray spectra \citep{Barret2024, Dupourqu2025, Barret2026} and to infer variability properties from dead time-affected light curves \citep{Huppenkothen2021}. In this work, we apply NPE to PPM.

Rather than an amortized approach, we adopt the truncated, sequential variant, which trains over multiple rounds, each time concentrating on the regions of parameter space most consistent with the target observation\footnote{By observation we mean all observations of a given source combined into a single pulse profile. This is possible because rotation-powered millisecond pulsars are highly stable in time.}. This significantly reduces the required number of training simulations compared to a single-round approach \citep[see e.g.,][]{Papamakarios2016-NPE, Lueckmann2017-SNPE, Greenberg2019-APT, Lueckmann2021}. The trade-off is the loss of amortization: the resulting network applies to one observation rather than to any observation drawn from the prior. In practice this loss is small, since only a handful of neutron star data sets are suitable for PPM. Amortization would matter more in situations where repeated inference is needed across many data sets, such as extensive simulation-based calibration or robustness checks of the underlying physics model. 

A further benefit of NPE is that it only requires a forward simulator, not an explicit likelihood function. This makes it possible to incorporate models that include stochastic processes that can be simulated but not written down in closed form. Although the likelihood was available in previous work, this capability may become important in the future for the next-generation X-ray telescopes and for more realistic models including stochastic processes, such as turbulent accretion flow onto accreting millisecond pulsars \citep{Das2025}. There, the turbulent accretion flow gives rise to changing hot spot shapes and variable surface temperature profiles, all of which cause the pulse profile to vary, producing a signal whose likelihood is intractable: a scenario that NPE can handle but nested sampling cannot.  

In this paper, we take a first step toward applying SBI to neutron star X-ray pulse profiles. Using TSNPE, we perform parameter inference on simulated pulse profiles with the aim of building and testing the pipeline. We verify that it recovers the injected parameters, demonstrate a speed-up relative to nested sampling, specifically MultiNest, and compare the trade-offs between the two approaches. We repeat the inference across a range of background levels relative to the pulsar signal to confirm that the method remains robust as the background level is altered (emulating the conditions expected for different telescopes). Together, this lays the groundwork for future applications to real data and to models that lie beyond the reach of likelihood-based inference. For cases in which the likelihood is available, we additionally show how neural importance sampling can be used to estimate the evidence and thereby enable model comparison \citep{Dax2023-IS}.  

The remainder of this paper is organized as follows. Section~\ref{sec:ppm_simulation} introduces the X-PSI forward model, the two hot region configurations we analyze, and the generation of the training simulations. In Section~\ref{sec:inference} we describe the inference: first the MultiNest baseline, then TSNPE and the associated data compression, network setup, posterior validation, and evidence estimation. Section~\ref{sec:results} compares the two methods in terms of posteriors, coverage, Bayesian evidence, and computational cost at 20\% background, and checks the robustness of the SBI posteriors across varying background levels. We discuss the limitations of the current approach, directions for future work, and our conclusions in Section~\ref{sec:discussion}.

\section{Pulse profile modeling and simulation}\label{sec:ppm_simulation}
In Section~\ref{subsec:xpsi} we describe the X-PSI PPM analysis framework, in Section~\ref{subsec:models} we introduce the neutron star models we analyze together with their priors, and in Section~\ref{subsec:simulations} we describe how the training data are simulated. 

\subsection{Pulse profile modeling using X-PSI: the forward model}\label{subsec:xpsi}
We use the open-source X-PSI package, versions \verb|v3.0.0| to \verb|v3.3.0| \citep{xpsi}, to generate synthetic pulse profiles. As in previous work \citep[see e.g.,][]{Riley2021-J0740, Salmi2022-J0740, Salmi2024-J0740, Choudhury2024-J0437}, X-PSI adopts the `Oblate Schwarzschild + Doppler' approximation to model the energy-resolved X-ray pulses from the neutron star \citep{Miller1998, Nath2002, Poutanen2003, Morsink2007, Lo2013, AlGendy2014, Bogdanov2019-part2, Ngo2026-error_oblate_schwarzschild}. This approximation accounts for the oblate shape of the neutron star and the relativistic effects due to its rapid spin, while treating the exterior spacetime as Schwarzschild. For a more detailed description of X-PSI, see \citet{Riley2019-J0030-xpsi, xpsi}.

We take NICER as the instrument, adopting the same \texttt{nicer\_20170601v003} redistribution matrix file (RMF) and \texttt{nicer\_20170601v005} ancillary response file (ARF) as used in real NICER analyses, with a synthesized exposure time of 1\,Ms and corresponding Poisson noise. The pulse profiles are generated over 32 phase bins and energy channels $[20, 200)$, corresponding to the photon energy range $0.2-2.0$ keV. We use the fully ionized hydrogen atmosphere model \texttt{NSX} \citep{Ho2001-atmosphere, Ho2009-atmosphere}. The emission is attenuated along the line of sight using the \texttt{tbnew}\footnote{\url{https://pulsar.sternwarte.uni-erlangen.de/wilms/research/tbabs/}} model for interstellar absorption \citep{Wilms2000-xrayISM}, parametrized by the neutral hydrogen column density $N_{\mathrm{H}}$. 

In contrast to real analyses, we adopt a simplified background model: a phase-independent (unpulsed) background with a power-law photon number spectrum, $N(E) \propto E^{-\Gamma}$ where $N(E)$ is the photon number flux and $\Gamma \in (1, 3)$ is the background photon index \citep[similar to the approach taken in previous NICER simulations in][]{Bogdanov2021-part3}. We fold the background spectrum through the instrument response and normalize it to a free parameter $N_{\rm bkg}$, the total expected background counts over the exposure, with a prior spanning 0 to 25\% of the total photon counts in the pulse profile (0 to $\sim1.8\times10^6$ background counts). Real analyses for NICER instead constrain the background either empirically -- using dedicated NICER background models such as 3C50 \citep{Remillard2022-3C50}, or imaging data from XMM-Newton or Chandra -- or by leaving the expected background counts free in each energy channel and marginalizing over them analytically \citep[see Section~3.4 of][for details]{Riley2019-J0030-xpsi}. Our power-law background model is therefore more restrictive than either approach.  

It is also a physical simplification, because the true background combines several different sources, each with their own spectral and temporal features. These include the other parts of the pulsar system such as a binary companion, nearby astronomical sources \citep[e.g. the active galactic nucleus near PSR~J0437--4715;][]{Bogdanov2019-part1}, diffuse emission in the field of view \citep[e.g. the pulsar wind nebula of PSR~J2124--3358;][]{Gonzalez-Caniulef2026-J2124}, cosmic X-ray background, and/or instrumental background, such as optical loading, energetic particles, and solar wind charge exchange. A single power law describes the non-thermal sky background reasonably well, but captures neither the spectral shape of the instrumental components nor any temporal or phase variability.

The impact of the simplified background depends on the instrument. For future telescopes such as eXTP and NewAthena, the predicted background is low and better constrained, so we expect the effect to be minor. However, for NICER the background is both high, exceeding our prior range for most pulsars, and composed of multiple astrophysical and particle components with distinct spectral shapes that a single power law cannot capture, so this case likely requires further work. See Section~\ref{sec:discussion} for further discussion on this topic.

\begin{deluxetable*}{llccc}
\tablecaption{Prior distributions and injected parameter values for the \texttt{ST} and \texttt{ST-U} models.\label{tab:priors}}
\tabletypesize{\footnotesize}
\tablewidth{\textwidth}
\tablehead{
  \colhead{Parameter} &
  \colhead{Description} &
  \colhead{Prior} &
  \colhead{Injected value, \texttt{ST}} &
  \colhead{Injected value, \texttt{ST-U}}
}
\startdata
$M$ [$M_{\odot}$]                            & gravitational mass                    & $U(1.0,3.0)$                     & 1.4     & 1.4  \\
$R_{\mathrm{eq}}$ [km]                       & coordinate equatorial radius          & $U(3r_g(1),16)$                  & 12.0    & 12.0 \\
$D$ [kpc]                                    & Earth distance                        & $U(0.5,2.0)$                     & 1.0     & 1.0  \\
$\cos(i)$                                    & cosine Earth inclination to spin axis & $U(0.0,1.0)$                     & 0.5     & 0.5  \\
$\phi_{p}$ [cycles]                          & $p$ region initial phase              & $U(0.0,1.0)$                     & 0.2     & 0.2  \\
$\Theta_{p}$ [radians]                       & $p$ region center colatitude          & \begin{tabular}[c]{@{}l@{}}$\cos\Theta_p \sim U(0.0,1.0) \text{ for ST}$ \\ $\cos\Theta_p \sim U(-1.0,1.0) \text{ for ST-U}$ \end{tabular}  & 1.22    & 1.22 \\
$\zeta_{p}$ [radians]                        & $p$ region angular radius             & $U(0.0,\pi/2)$                   & 0.75    & 0.75 \\
$\log_{10}\!\left(T_{p}\,[\mathrm{K}]\right)$& $p$ region effective temperature      & $U(5.1,6.8)$                     & 6.2     & 6.2  \\
$\phi_{s}$ [cycles]                          & $s$ region initial phase              & $U(0.0,1.0)$                     & \nodata & 0.1  \\
$\Theta_{s}$ [radians]                       & $s$ region center colatitude          & $\cos\Theta_s \sim U(-1.0,1.0)$  & \nodata & 2.0  \\
$\zeta_{s}$ [radians]                        & $s$ region angular radius             & $U(0.0,\pi/2)$                   & \nodata & 1.0  \\
$\log_{10}\!\left(T_{s}\,[\mathrm{K}]\right)$& $s$ region effective temperature      & $U(5.1,6.8)$                     & \nodata & 5.8  \\
$N_{\mathrm{H}}$ [$10^{20}$\,cm$^{-2}$]      & interstellar neutral H column density & $U(0.0,10.0)$                    & 2.0     & 2.0  \\
$N_{\rm bkg}$ [counts]                       & background counts                     & \begin{tabular}[c]{@{}l@{}}$U(0, 1.77\times10^6) \text{ for ST}$ \\ $U(0, 1.80\times10^6)\text{ for ST-U}$ \end{tabular}                                       & $1.33\times10^6$ & $1.35\times10^6$ \\ 
$\Gamma$                                     & background photon index               & $U(1, 3)$                        & 2.0     & 2.0      
\enddata
\tablecomments{Parameters of the primary hot region carry the subscript $p$; those of the secondary hot region carry the subscript $s$. Prior distributions and injected parameter values are identical between the two models except for the secondary hot region, which the \texttt{ST} model does not include, and for the primary hot region colatitude which is restricted to one hemisphere for \texttt{ST}. The injected $N_{\rm bkg}$ corresponds to 20\% of the total photon counts in the pulse profile, and the prior upper limit to 25\%, for both models; the small difference between the models is due to their different total photon counts.}
\end{deluxetable*}

\subsection{Hot region configurations}\label{subsec:models}
We examine two neutron star models in which the hot emitting regions are described by either one or two independent circular, uniform-temperature spots: the \texttt{ST} (\textit{Single Temperature}) and \texttt{ST-U} (\textit{Single Temperature-Unshared}) models, respectively. See Table~\ref{tab:priors} for the prior distributions used. Note that priors on a given parameter are identical for the \texttt{ST} and \texttt{ST-U} runs, except for the secondary hot spot which the \texttt{ST} model does not include, and for the primary hot region colatitude which is restricted to the Northern hemisphere for the \texttt{ST} model to break the reflection degeneracy. The prior for the background level, $N_{\rm bkg}$, is set to $\sim U(0, 1.8\times10^6)$ counts spanning 0 to 25\% of the total photon counts in the pulse profile. The upper limit is slightly higher for the \texttt{ST-U} model than for \texttt{ST}, because the second hot spot increases the total photon count. For more details on the X-PSI model naming convention and associated parameters, see \citet{Riley2019-J0030-xpsi, Vinciguerra2023-J0030-sim}. 

\subsection{Training simulations}\label{subsec:simulations}
Synthetic data are generated by first drawing a parameter vector $\theta$ from the prior (Table~\ref{tab:priors}). For each parameter vector, X-PSI computes the expected counts per channel and phase bin, and an independent Poisson realization of these expected counts is drawn and added, using a different random seed for each draw.

The inference in this paper focuses on specific injected values for each parameter, which are shown in Table~\ref{tab:priors}. The background is set at 20\% of the total photon counts in the pulse profile, which corresponds to $N_{\rm bkg}=1.33\times10^{6}$ counts for \texttt{ST} and $1.35\times10^{6}$ counts for \texttt{ST-U}. As mentioned before, the two values differ slightly because the models have different total photon counts. To assess the influence of the background level, we repeat the inference at 1\% and 10\% background ($N_{\rm bkg}=5.37\times10^{4}$ and $5.91\times10^{5}$ counts for \texttt{ST}; $5.47\times10^{4}$ and $6.02\times10^{5}$ counts for \texttt{ST-U}).

These percentages of background contribution to the total number of counts are chosen to roughly represent current and future situations. A $\sim$10--20\% background level corresponds to that detected in the NICER observations of a bright millisecond pulsar\footnote{The NICER background level is inferred indirectly, and its contribution can vary between different analyses.}, such as PSR~J0437--4715 \citep{Stammler2026-J0437}, or to that expected in the eXTP observations of a faint pulsar, such as PSR~J0614--3329.  With the NewAthena Wide-Field Imager, observations of such a pulsar would instead give a background level of $\sim$ 1\% or less (Guillot et al. 2026, to be submitted). 

\section{Inference}\label{sec:inference}
In Section~\ref{subsec:baseline_MN} we briefly summarize the standard PPM pipeline, in which inference is performed with the nested sampler MultiNest and which we take as our baseline for the SBI comparison. In Sections~\ref{subsec:sbi} and \ref{subsec:tsnpe} we explain the inference with the SBI method TSNPE. In Section~\ref{subsec:pca} we describe the dimensionality reduction applied to the training data before it is fed to the neural network, and in Section~\ref{subsec:setup} we summarize the technical setup of the TSNPE. In Section~\ref{subsec:SBCC} we show how the quality of the SBI posterior is validated, and finally in Section~\ref{subsec:IS} we describe how the evidence is estimated. Complete information for each inference run, including the exact X-PSI and SBI versions, data products, posterior sample files, and all the analysis files are available in a Zenodo repository\footnote{\url{https://doi.org/10.5281/zenodo.23312851}} \citep{Hoogkamer2026-Zenodo}.

\subsection{Baseline analysis with MultiNest}\label{subsec:baseline_MN}
In the standard PPM workflow, X-PSI is coupled to the nested sampling algorithm MultiNest \citep{Feroz2008-MultimodalNestedSampling, Feroz2009-Multinest, Feroz2019-Multinest} via PyMultiNest \citep{Buchner2016-PyMultiNest} to compute posterior samples and evaluate the Bayesian evidence for model comparison. We use this workflow as the baseline against which the SBI results are compared.

We run MultiNest with 1000 live points, a sampling efficiency of 0.3, and an evidence tolerance of 0.1. These settings are comparable to what has been used in exploratory runs in previous NICER analysis\footnote{We note that changes in the posterior between exploratory and headline results are usually minor.} \citep{Riley2019-J0030-xpsi, Vinciguerra2023-J0030-sim}, but are less computationally demanding than those adopted for headline results \citep[see e.g.][]{Choudhury2024-J0437, Salmi2024-J0740}. Importance nested sampling is used to improve the accuracy of the evidence calculation without changing the way MultiNest explores the parameter space\footnote{By default, MultiNest calculates the evidence from only the samples that satisfy the likelihood constraint $\mathcal{L}>\mathcal{L}_i$. Importance nested sampling uses all the points, including the rejected ones, assigning each an importance weight, which is the ratio of the prior to the sampling density (see Section~4.1 of \citealt{Feroz2019-Multinest} for a detailed description)} \citep{Feroz2019-Multinest}. 

\subsection{Simulation-based inference: neural posterior estimation}\label{subsec:sbi}
Bayesian inference aims to compute the posterior distribution $p(\theta|x_0)$, the probability of a set of parameters $\theta$ given observed data $x_0$. This requires a prior $p(\theta)$ and explicit evaluations of the likelihood $p(x|\theta)$, which can be computationally prohibitive when the likelihood is expensive to evaluate or intractable.

SBI bypasses the need for explicit likelihood evaluations by learning the posterior directly from simulated data \citep{Cranmer2020-SBI, Deistler2025-sbi}. There are three main approaches: neural posterior estimation \citep[NPE;][]{Papamakarios2016-NPE, Greenberg2019-APT}, neural likelihood estimation \citep[NLE;][]{Papamakarios2019-NLE}, and neural ratio estimation \citep[NRE;][]{Hermans2020-NRE}.

This paper focuses on NPE, which works as follows: first, a simulator is used to generate a data set of parameter--simulation pairs $\{(\theta_i, x_i)\}$. A neural density estimator $q_\phi(\theta|x)$, typically a normalizing flow \citep{dinh2014,Papamakarios2021-NF}, is then trained to learn the probabilistic mapping from simulated data to parameters. After training, the network is evaluated at the observed data $x_0$ to obtain the posterior for that particular data set. The main computational cost lies in generating the training data and in training the normalizing flow; once the network is trained, inference reduces to a forward pass and can be performed in seconds. 

\begin{figure*}[!]
\centering
\includegraphics[width=1.\textwidth]{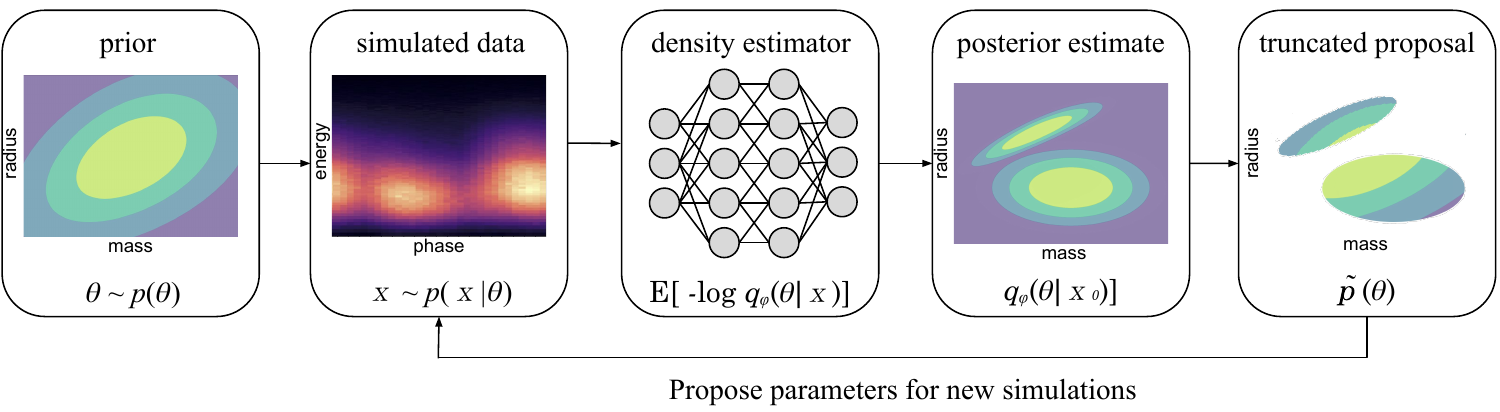}  
\caption{Schematic of TSNPE. In the first round, parameters are sampled from the prior (here mass and radius are shown as an example). The simulator is then run to produce pulse profiles, and the neural density estimator is trained to approximate the posterior. In subsequent rounds, parameters are sampled from the prior but rejected if they lie outside the support of the current approximate posterior, also known as the truncated proposal. Figure adapted from \citet{Deistler2022-TSNPE}.}
\label{fig:tsnpe}
\end{figure*}

NPE can be applied in either an amortized or a sequential framework. In the amortized setting the network is trained once on simulations spanning the full prior volume, after which it can be applied to any observation for which that prior is appropriate without retraining\footnote{This holds only as long as the modeling does not change. If, for example, the instrument response changes, the simulator changes too, and retraining is needed.}. In the sequential setting (SNPE), training is iterated over several rounds, each concentrating simulations more tightly on the region of parameter space consistent with a specific observation $x_0$, which substantially reduces the number of simulations needed for an accurate posterior \citep[see e.g.,][]{Papamakarios2016-NPE, Lueckmann2017-SNPE, Greenberg2019-APT, Lueckmann2021}. 

Since the pulse profiles of rotation-powered millisecond pulsars are highly stable in time, all observations of a given source are combined into a single profile, so each source yields one data set rather than many. Only a handful of neutron stars are bright enough for NICER to accumulate the photon counts that pulse profile modeling requires, and while NewAthena and eXTP will improve the data quality, they are not expected to substantially increase the number of feasible targets. Amortization across many observations therefore offers little benefit here, whereas simulation efficiency does. We therefore adopt the sequential approach. 

\subsection{Truncated sequential neural posterior estimation}\label{subsec:tsnpe}
In this work, we use TSNPE \citep{Deistler2022-TSNPE}, which follows the sequential NPE formalism but uses a proposal that is a truncated version of the prior. In the first round, where no (intermediate) posterior estimate is yet available, parameters are drawn directly from the prior. In each subsequent round, parameters are again drawn from the prior but rejected before simulation if they lie outside the support of the current approximate posterior. This distribution is referred to as the truncated proposal, $\tilde{p}(\theta)$ (see Figure~\ref{fig:tsnpe}). Because this proposal is proportional to the prior within its support, sampling from it and minimizing the standard NPE loss requires no correction term, so the same loss, $\mathcal{L}$, can be used every round: 
\begin{equation}\label{eq:loss}
    \underset{\phi}{\text{min}}\,\mathcal{L} = \underset{\phi}{\text{min}}~\mathbb{E}_{\theta \sim \tilde{p}(\theta)} \mathbb{E}_{x\sim p(x|\theta)}[-\log q_\phi (\theta|x)],
\end{equation} 
which, for a sufficiently expressive density estimator, is minimized if and only if $q_\phi(\theta|x)=p(\theta|x)$ for all $x\in\text{supp}(p(x))$ \citep{Paige2016, Papamakarios2016-NPE}. 

Truncating the prior in this way has two advantages. First, it stabilizes training: standard sequential NPE methods require a modification to the loss to correct for the use of non-prior proposals. This modification can introduce instabilities such as the leakage of posterior mass beyond the prior bounds. Training with an unmodified maximum likelihood loss avoids this \citep{Deistler2022-TSNPE}. Second, it prevents the normalizing flow from placing posterior mass in regions excluded by the prior: flows have unbounded support and can assign probability to parameter values outside the prior bounds, but rejecting candidates before simulation avoids this. 

The support of the truncated proposal is the highest probability region (HPR) of the current posterior estimate, defined as the smallest region containing $1-\eta$ of the total posterior mass (within it the proposal is proportional to the prior). We adopt $\eta=10^{-3}$, so that at most $0.1\%$ of the posterior mass is excluded per round.

Each training round consists of the following steps: 
\begin{enumerate}
    \item Draw parameters from the current proposal, $\theta_i \sim \tilde{p}(\theta)$, which equals the prior $p(\theta)$ in the first round. 
    \item Generate synthetic data $x_i \sim p(x|\theta_i)$ using the simulator described in Section~\ref{sec:ppm_simulation}. 
    \item Train the neural density estimator $q_\phi(\theta|x)$ on all parameter--simulation pairs $\{(\theta_i, x_i)\}$ accumulated so far, including those from previous rounds.
    \item Evaluate the trained estimator at the observation to obtain $q_\phi(\theta|x_0)$, and update the proposal by truncating the prior to the HPR of the current posterior estimate.
\end{enumerate}
Successive rounds thus concentrate the simulation budget on the region of parameter space compatible with $x_0$, which is what makes the method simulation-efficient relative to a single-round (amortized) approach. The training continues until the density estimator converges to a stable posterior distribution (see Section~\ref{subsec:SBCC} for how the quality of the posterior is assessed). 

To sample the truncated proposal and generate samples for the next training round, rejection sampling or sampling importance resampling \citep[SIR;][]{Rubin1988-SIR, Smith1992-SIR} can be used. Rejection sampling becomes inefficient once the HPR occupies a small fraction of the prior volume, since almost every draw is rejected. We therefore use SIR instead, which uses the approximate posterior itself as the sampling distribution. To obtain one sample from the truncated proposal, SIR draws $K$ candidates from the approximate posterior $\theta_1, \dots, \theta_K \sim q_\phi(\theta|x_0)$, assigns each an importance weight $w_i = \frac{p(\theta_i)}{q_\phi(\theta_i|x_0)}$, normalizes the weights so that they sum to one, and returns a single $\theta_i$ drawn from the categorical distribution defined by them. The weights correct for the fact that candidates are proposed from the approximate posterior rather than the prior, so that the retained samples follow the truncated prior in the limit of large $K$. Because $q_\phi(\theta|x_0)$ already concentrates on the region of interest, few candidates are wasted. Following \citet{Deistler2022-TSNPE}, we use a high value, $K = 1000$.

\subsection{Dimensionality reduction via principal component analysis}\label{subsec:pca}
A key challenge in SBI is the often necessary compression of the high-dimensional data (here: the pulse profile) into low-dimensional summary statistics that are informative with respect to the parameters of interest. This compression isolates the informative components, discards noise and redundant features, and reduces the computational cost of the inference \citep{Cranmer2020-SBI}. In general, the summary statistics should retain as much information about the parameters $\theta$ as possible, such that the posterior derived from them approximates the true posterior. 

In this work we use principal component analysis \citep[PCA;][]{Jolliffe2016-PCA} to reduce the dimensionality of the data, as in \citet{Parker2022-AGNnn} and \citet{Barret2024}. PCA performs a linear dimensionality reduction: using a singular value decomposition of the simulations generated in the first round of the sequential NPE process, it identifies a set of mutually orthogonal directions along which the data vary most, ranked by the variance they capture, and projects the data onto these principal components.

We first square-root transform the pulse profiles, which stabilizes the variance of Poisson-distributed data \citep{Anscombe1948-transform} and prevents the high-count channels from dominating the decomposition. Each pulse profile is then reduced from $180\times32=5760$ phase--channel bins to 60 principal components, retaining $99.996\%$ of the variance. The PCA is fit once on the first-round simulations and held fixed thereafter, so that the summary statistics remain consistent across rounds and earlier simulations stay usable as training data.

\subsection{Deep learning setup}\label{subsec:setup}
As our neural density estimator we use a normalizing flow \citep[see, e.g.,][]{dinh2014, Papamakarios2019, Kobyzev2019}, specifically a coupling neural spline flow \citep{Durkan2019, Muller2018}. A normalizing flow approximates the target distribution by transforming a simple, analytically known base distribution -- here a standard normal $\mathcal{N}(0, 1)$ -- into a more complex one through a sequence of invertible (bijective) and differentiable transformations. In our case these are monotonic rational-quadratic spline bijections applied in a coupling architecture, each parametrized by neural networks to allow for greater flexibility. The resulting model allows both sampling from and evaluation of the log-density of the approximated distribution.

The normalizing flow is built from 12 spline transformations for the \texttt{ST} model and 15 for the \texttt{ST-U} model. In both cases, each transformation is conditioned by a residual network of two residual blocks with 128 hidden units per layer. The flow is trained by minimizing the negative log-likelihood (see Equation~\ref{eq:loss}) with the Adam optimizer \citep{Kingma2017}, using a batch size of 256, a learning rate of $1\times10^{-4}$, and 20\% of the samples reserved for validation. Per round, training stops early once the validation loss has not improved for 20 consecutive epochs, up to a maximum of 500 epochs, which on average happens after $\sim200$ epochs. 

We run TSNPE for several rounds, stopping once the approximate posterior has stabilized between successive rounds: the credible intervals shrink substantially in the first rounds and then level off. This point is reached after 4 rounds for the \texttt{ST} model and 5 rounds for the \texttt{ST-U} model. In the final round, instead of a single flow, we train a deep ensemble of five normalizing flows on the same simulations, each initialized with a different random seed \citep{Lakshminarayanan2017-ensemble}. We take the ensemble posterior to be the average of the five normalizing flows, which reduces the variance of the density estimate and yields a more robust posterior.

All inference in this work is carried out with the \texttt{sbi} package \citep{BoeltsDeistler2025-sbi}, version \texttt{0.25.0}. 

\subsection{Validation of the approximate posterior}\label{subsec:SBCC}
For the approximate posterior $q_\phi(\theta|x_0)$ to converge to the true posterior $p(\theta|x_0)$, TSNPE requires the estimated posterior to be broader than the true one. An overconfident estimate would exclude regions of parameter space that actually carry posterior mass. Because truncation of the proposal is applied before simulation, such regions are lost permanently in all subsequent training rounds. To check whether the posterior is, on average, sufficiently broad, we apply the expected coverage test \citep[see, e.g.,][]{Talts2018-ValidationPP, Hermans2022-crisis, Deistler2022-TSNPE}.

The expected coverage is obtained by averaging the coverage over many pairs $(\theta^*, x^*)$, where $\theta^*$ is drawn from the truncated proposal and $x^* \sim p(x \mid \theta^*)$ is the corresponding simulated pulse profile. For each nominal credible level $1-\alpha$, we compute the fraction of pairs for which $\theta^*$ falls inside the $1-\alpha$ credible region of $q_\phi(\theta|x^*)$. For a well-calibrated posterior, the empirical coverage should match the nominal level at all credible levels $(1-\alpha) \in [0,1]$, i.e., lie on the diagonal. For TSNPE it is especially important that the posterior is not overconfident at high credible levels, that is, that the empirical coverage matches or slightly exceeds the nominal level as $\alpha \rightarrow 0$. Overconfidence in this regime means that the ground-truth parameters $\theta^*$ are being excluded from the highest probability region used to define the proposal for the next round.

Because the TSNPE posterior is a normalizing flow that can be both sampled and evaluated in closed form, this diagnostic is cheap and scalable, and can be run after every training round without additional simulations or retraining. 

\subsection{Evidence calculation via neural importance sampling}\label{subsec:IS}
Model comparison is an important aspect of pulse profile modeling. Because we retain access to the likelihood, we can estimate the Bayesian evidence within our SBI framework directly with importance sampling (IS), following the approach of \citet{Dax2023-IS} -- often referred to as ``neural importance sampling'' when the approximate posterior is a neural density estimator \citep{Muller2018}.

The evidence is estimated by drawing $n$ samples from a proposal, here the approximate posterior which is the trained normalizing flow, $\theta_i \sim q_\phi(\theta|x_0)$, and assigning to each an importance weight $w_i=p(x_0|\theta_i)\,p(\theta_i)/q_\phi(\theta_i|x_0)$, i.e., the ratio of the unnormalized posterior (likelihood $\times$ prior) to the proposal density. Evaluating the weights thus requires computing the likelihood at each of the $n$ samples. For a perfect proposal the weights are constant; otherwise their spread quantifies the mismatch between proposal and posterior, summarized by the effective sample size $n_{\mathrm{eff}}=\left(\sum_i w_i\right)^2/\sum_i w_i^2$. The corresponding sample efficiency $\epsilon=n_{\mathrm{eff}}/n \in(0,1]$ measures the quality of the proposal, with $\epsilon=1$ for a perfect proposal and $\epsilon \to 0$ for a poor one.

The evidence is the normalizing constant of the unnormalized posterior. Because the proposal $q_\phi(\theta|x_0)$ integrates to one, the evidence can be written as the expectation of the importance weights over parameters drawn from the proposal. We therefore estimate the evidence by the sample mean of the weights, $p(x_0)=\frac{1}{n}\sum_i w_i$, which is unbiased for any proposal whose support covers that of the posterior. The standard deviation of the log evidence is given by $\sigma_{\log p(x_0)}=\sqrt{(1-\epsilon)/(n \cdot \epsilon)}$. In this work we set $n=5\times10^4$, following \citet{Dax2023-IS}.

\section{Results}\label{sec:results}
We now compare SBI with the MultiNest baseline for both the \texttt{ST} and \texttt{ST-U} models. Section~\ref{subsec:posterior_validation} validates the SBI posteriors, Section~\ref{subsec:evidence} compares the Bayesian evidences obtained by the two methods, and Section~\ref{subsec:cost} quantifies and compares their computational cost. 

\subsection{Posterior validation}\label{subsec:posterior_validation}
\begin{figure*}[!]
\centering
\includegraphics[width=0.8\textwidth]{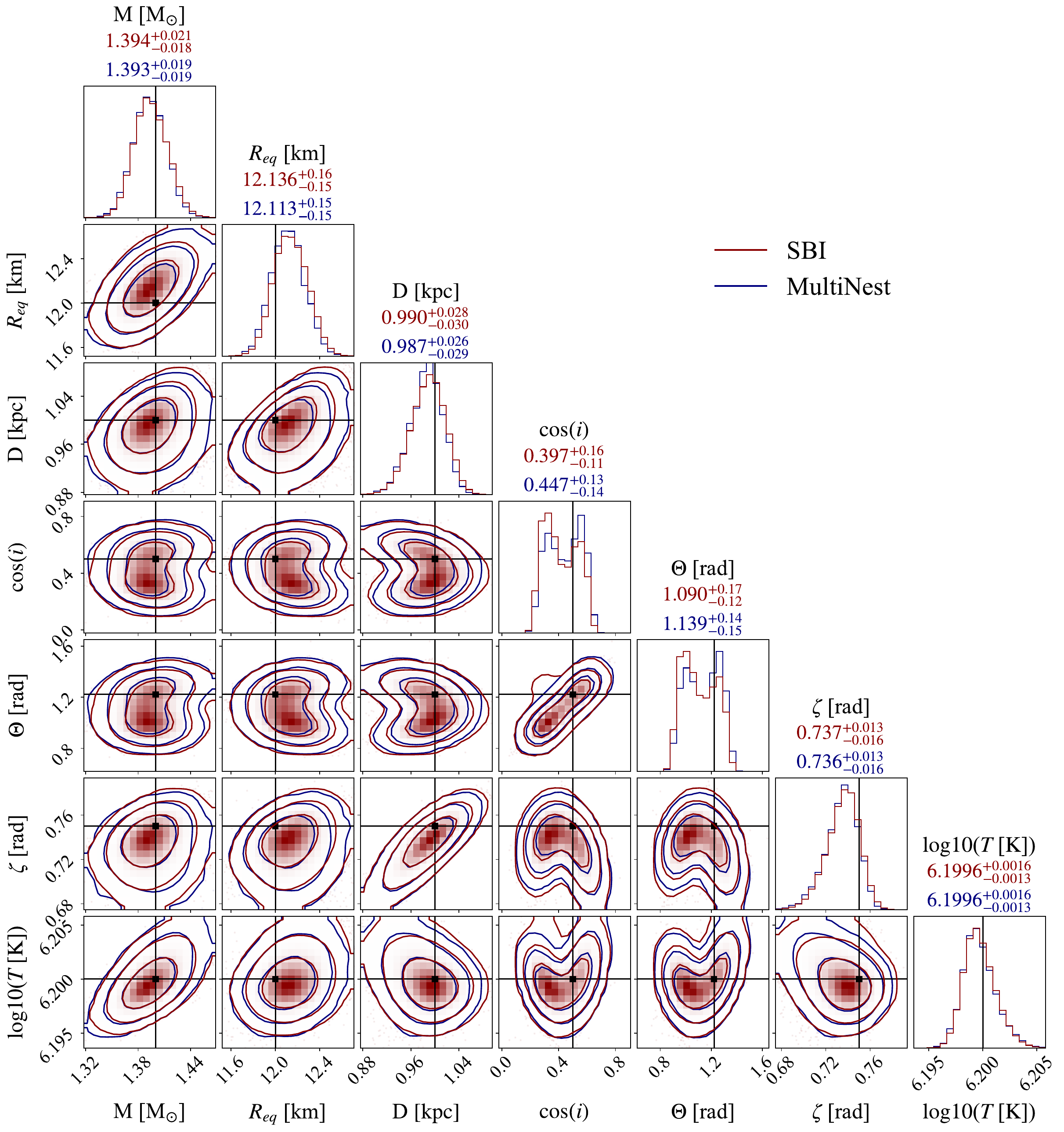}  
\caption{Posterior distributions for some parameters of the \texttt{ST} model at 20\% background, comparing SBI (dark red) with MultiNest (navy). Diagonal panels show the one-dimensional marginal distributions. The contours in the off-diagonal panels contain the 68.3\%, 95.4\%, and 99.7\% credible regions for the two-dimensional joint distributions. Black solid lines mark the injected parameter values (see Table~\ref{tab:priors}). Titles report the median and gray credible interval for each parameter.}
\label{fig:corner_ST_20bkg_small}
\end{figure*}

First, we verify that SBI recovers the injected parameter values of the synthetic data with credible intervals consistent with those obtained from nested sampling. Figure~\ref{fig:corner_ST_20bkg_small} shows the posterior distributions of some of the parameters of the \texttt{ST} for a single observation with 20\% background (injected values listed in Table~\ref{tab:priors}). The full posteriors for all parameters of the \texttt{ST} and \texttt{ST-U} models with 20\% background are shown in Figures~\ref{fig:corner_ST_20bkg} and~\ref{fig:corner_STU_20bkg} in the Appendix, respectively. We compare the baseline MultiNest run, using $1\times10^3$ live points and a sampling efficiency of 0.3, with SBI, run for 4 rounds with $20\,000$ simulations each for \texttt{ST} and 5 rounds of $25\,000$ simulations each for \texttt{ST-U}; in the final round, an ensemble of five normalizing flows is trained (see Section~\ref{sec:inference} for details). For both models, MultiNest and SBI recover the injected parameters and obtain overlapping credible intervals at the 68.3\% level. 

\begin{figure*}[!]
\centering
\includegraphics[width=.9\textwidth]{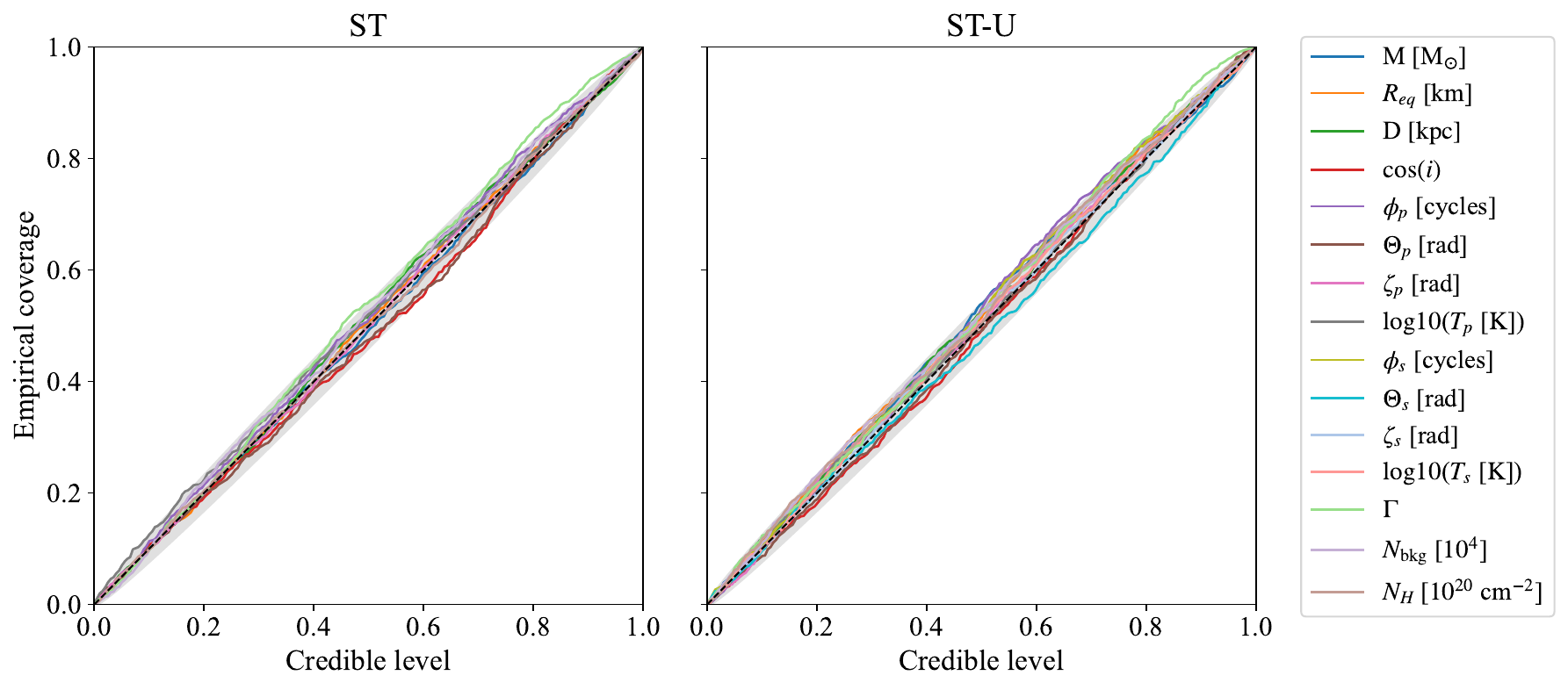}  
\caption{Empirical coverage of the SBI posteriors at 20\% background for the \texttt{ST} (left) and \texttt{ST-U} (right) model parameters, using 1000 simulations. For each simulation and its corresponding parameter vector, the credible level at which the injected parameter value is recovered is computed from the marginal posterior. The curves show the fraction of simulations for which the truth falls within a given credible interval as a function of that interval's nominal credible level. The diagonal corresponds to perfect calibration; curves lying above (below) it indicate conservative (overconfident) posteriors. The gray band is the $99\%$ confidence region for a calibrated posterior.}
\label{fig:coverage_ST_STU}
\end{figure*}

Second, we test the empirical coverage of SBI because agreement on a single data set does not establish that the posteriors are statistically well-behaved in general. Figure~\ref{fig:coverage_ST_STU} shows the empirical coverage of the posteriors, computed for 1000 simulations (see Section~\ref{subsec:SBCC} for the methodology). For the \texttt{ST} model, the inclination $\cos(i)$ and the primary hot region colatitude $\Theta_p$ are mildly overconfident, dipping below the gray band which is the $99\%$ confidence region for a calibrated posterior. All remaining parameters in both models fall within the gray band or sit slightly on the conservative side, indicating well-calibrated to conservative posteriors. 

\begin{figure*}
    \centering
    \includegraphics[width=\textwidth]{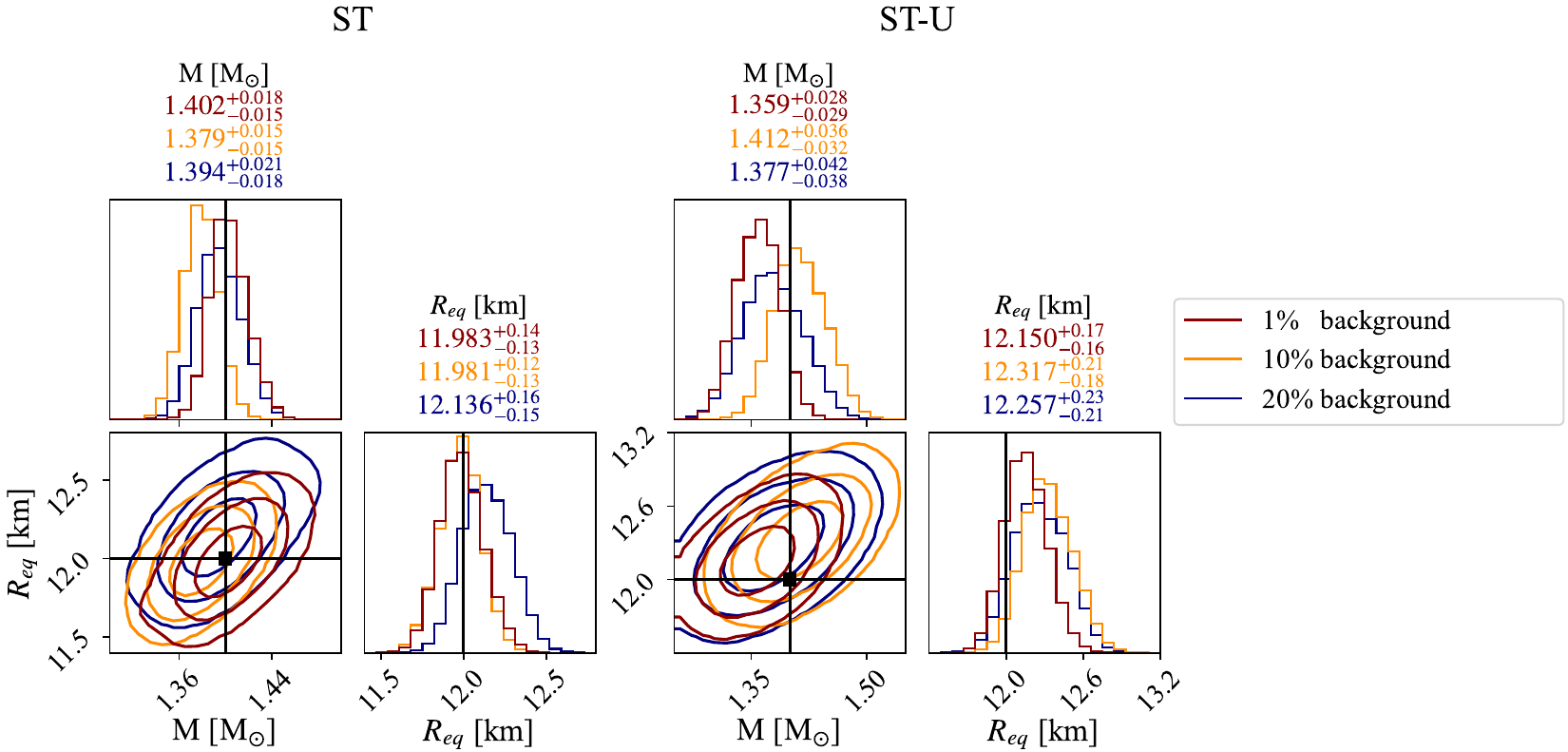}\\[1ex]
    \includegraphics[width=\textwidth]{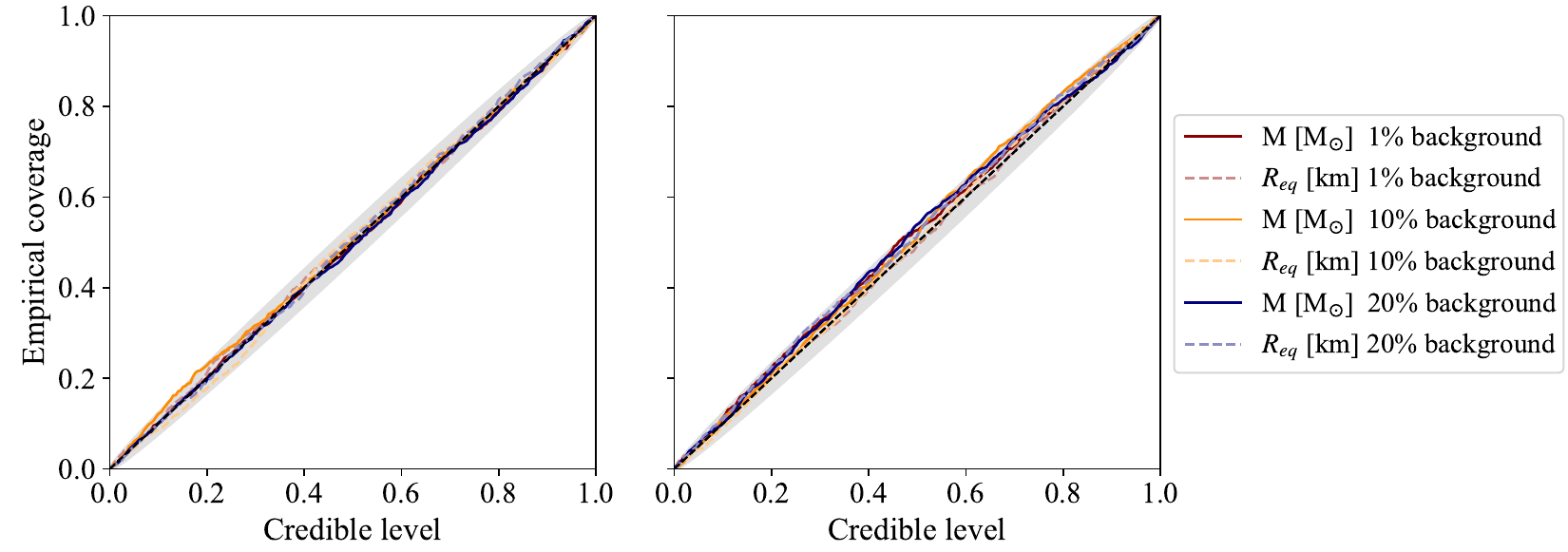}
    \caption{SBI posteriors for the \texttt{ST} (left) and \texttt{ST-U} (right) models at three background levels: 1\% (dark red), 10\% (orange), and 20\% (navy). Top: posterior distributions of the mass $M$ and equatorial radius $R_{eq}$; black lines mark the injected values. Bottom: empirical coverage of the same posteriors for $M$ (solid) and $R_{eq}$ (dashed), where the diagonal indicates perfect calibration. See Figures~\ref{fig:corner_ST_20bkg_small} and \ref{fig:coverage_ST_STU} for more details about the figure elements.}
    \label{fig:corner_coverage_ST_STU_bkg_lvls}
\end{figure*}

To confirm that SBI remains reliable across background levels, the top panels of Figure~\ref{fig:corner_coverage_ST_STU_bkg_lvls} show the mass--radius posteriors for the \texttt{ST} and \texttt{ST-U} models at 1\%, 10\%, and 20\% background (the full posteriors for all parameters at 1\% background are shown in Figures~\ref{fig:corner_ST_1bkg} and~\ref{fig:corner_STU_1bkg} in the Appendix). In every case, SBI recovers the injected mass--radius values within the 95\% credible interval, and in most cases within the 68.3\% interval. For the \texttt{ST-U} model, the radius is biased slightly high at all three background levels, with the injected value lying within the 68.3\% level at 1\% background, but just outside it at 10\% and 20\%. However, this is likely caused by the randomness of drawing a single observation, not a bias in SBI, as MultiNest shows the same trend. The empirical coverage test in the bottom panels of Figure~\ref{fig:corner_coverage_ST_STU_bkg_lvls} supports this conclusion: the mass and radius fall within the gray band for both models at all three background levels, indicating well-calibrated posteriors.   

\subsection{Bayesian evidence}\label{subsec:evidence}
Table~\ref{tab:evidence} compares the log-evidence, $\ln\mathcal{Z}$, of the \texttt{ST} and \texttt{ST-U} models at the 20\% background obtained with two different methods: importance nested sampling with MultiNest, and neural importance sampling using $n=5\times10^{4}$ samples from the (unnormalized) neural posterior of SBI (see  Section~\ref{subsec:IS}). For both models the two evidence estimates are similar to within $\sim0.2-0.4$ in $\ln\mathcal{Z}$. Following \citet{Kass1995}, a difference of less than $1.0$ in $\ln\mathcal{Z}$ is ``not worth more than a bare mention''. The agreement between the two methods is therefore well within this threshold, and the error in the evidence is unlikely to affect the scientific conclusions. 

The neural importance sample efficiencies $\epsilon=n_{\mathrm{eff}}/n$ are $3.09\%$ and $0.40\%$ for \texttt{ST} and \texttt{ST-U} models, respectively. Such low sample efficiencies do not bias the evidence estimate, but do increase its variance \citep{Dax2023-IS}. To verify that the estimate has converged, we plot the evidence estimate as a function of the number of importance samples $n$. Figure~\ref{fig:evidence_convergence} in the Appendix shows $\ln\mathcal{Z}(n)$ reaching a stable plateau, within the uncertainties, for both models confirming the convergence. 

\begin{deluxetable}{lcc}
\tablecaption{Log-evidence estimates from MultiNest (importance nested sampling) and SBI (neural importance sampling with $n=5\times10^{4}$). Log-evidences are relative to $\ln\mathcal{Z}_{0} = -27\,033$ (\texttt{ST}) and $-27\,041$ (\texttt{ST-U}). The MultiNest runs use exploratory sampler settings: $10^{3}$ live points and a sampling efficiency of $0.3$.\label{tab:evidence}}
\tablewidth{\columnwidth}
\tablehead{
  \colhead{} &
  \colhead{\texttt{ST} model} &
  \colhead{\texttt{ST-U} model}
}
\startdata
$\ln\mathcal{Z}$ (MultiNest)            & $0.03 \pm 0.36$       & $0.32 \pm 0.21$ \\
$\ln\mathcal{Z}$ (SBI)                  & $0.40 \pm 0.03$       & $0.51 \pm 0.07$ \\
$\Delta\ln\mathcal{Z} $ (MultiNest $-$ SBI)& $0.37 \pm 0.37$    & $0.19 \pm 0.22$  \\
Sample efficiency $\epsilon$ (SBI)      & $3.09\%$              & $0.40\%$ \\
\enddata
\end{deluxetable}

Important to note is that for the evidence calculation with MultiNest we enabled importance nested sampling, whereas most NICER analyses instead reach comparable evidence precision through higher sampler settings (with `vanilla' nested sampling). To check that this choice does not drive the agreement, we repeated the MultiNest runs without importance nested sampling, increasing the sampler settings until the evidence stabilized. The \texttt{ST} model required $4\times10^{3}$ live points at a sampling efficiency of $0.01$ giving $\ln\mathcal{Z} = -27\,033.01 \pm 0.12$ which is comparable to the values obtained with both SBI and our baseline (importance nested sampling) MultiNest run.

For the \texttt{ST-U} model, the evidence decreases towards the SBI evidence estimate as the sampler settings are increased: at $4\times10^{3}$ live points and a sampling efficiency of $0.01$, we obtain $\ln\mathcal{Z} = -27\,036.79 \pm 0.12$. This still differs from both the SBI value and the baseline MultiNest value by $4.72 \pm 0.14$ in $\ln\mathcal{Z}$. However, limited computational resources prevented us from testing higher sampler settings. Given the clear trend of $\ln\mathcal{Z}(n_{\rm live})$ decreasing toward the SBI value with increasing sampler settings, we consider it likely that full convergence would bring the two into agreement.
 
All in all, the agreement with two independent MultiNest configurations for the \texttt{ST} model, and the trend of $\ln\mathcal{Z}(n)$ toward the SBI value for the \texttt{ST-U} model, support the reliability of the SBI evidence in both cases. 

\subsection{Computational cost}\label{subsec:cost}
To compare the computational cost of the two methods at 20\% background, we consider two metrics: the number of forward model calls -- likelihood evaluations for MultiNest versus simulations for SBI -- and the total inference time. Neither comparison is entirely fair. MultiNest's likelihood evaluation count also includes parameter vectors rejected by the prior before any forward model is run, and for the vectors that do pass, each evaluations additionally computes the likelihood itself, which SBI never does. The two methods were also run on different hardware (see below). Together, however, they give a decent indication of the relative cost.

We first compare the number of forward model calls, a standard metric in the SBI literature \citep[e.g.,][]{Dax2023-IS, Dupourqu2025} because it is hardware-agnostic: X-PSI is currently CPU-bound, whereas SBI simulates on a CPU but trains on a GPU. By comparing the number of forward model calls, the comparison also remains valid for future GPU-accelerated versions of X-PSI. This assumes that calculations of the physical model simulation incur the dominant cost in either approach, and that the computational cost for either training the normalizing flow (SBI) or for other nested sampling operations is negligible. As mentioned above, the likelihood evaluation count includes parameter vectors rejected by the prior before the forward model is run: for the \texttt{ST-U} model, for instance, the two hot regions are not allowed to overlap, so vectors that violate this constraint are rejected yet still counted as likelihood evaluations.

As shown in Table~\ref{tab:cost}, for the \texttt{ST} model MultiNest needs $2.6\times10^6$ likelihood evaluations, whereas SBI needs $8\times10^4$ simulations to get similar posteriors, which is a factor of $32.7$ fewer forward model calls. For the \texttt{ST-U} model, MultiNest uses $3.0\times10^7$ likelihood evaluations versus $1.25\times10^5$ simulations for SBI, which is a factor of $241.8$ fewer. 

We next compare the total inference time. MultiNest required $\sim296.4$ core hours for \texttt{ST} and $\sim6207.2$ core hours for \texttt{ST-U} (dual AMD EPYC 7H12 CPUs). SBI, in contrast, took $\sim3.3$ core hours on a CPU (Intel Core i9-13900H) to generate the pulse profiles plus $\sim2.1$ hours to train the density estimator on a GPU (NVIDIA RTX 4060) for \texttt{ST}, totaling $\sim5.4$ hours. For \texttt{ST-U} it took $\sim10.1$ CPU core hours plus $\sim4.0$ GPU hours, totaling $\sim14.1$ hours. This corresponds to a speed-up of $54.9$ and $440.2$ for \texttt{ST} and \texttt{ST-U}, respectively. 

Across both metrics, the speed-up is larger for the more complex model, so the advantage of TSNPE over MultiNest becomes more prominent as model complexity grows. Note that these are conservative estimates of the true speed-up. The number of likelihood evaluations for MultiNest mentioned here are for runs using exploratory sampler settings, whereas our headline results use higher sampler settings to ensure convergence and therefore require more likelihood evaluations. For example, measured instead against a converged \texttt{ST} run (without importance nested sampling) described above, the speed-up increases from $32.7$ to a factor of $210.1$ when comparing the number of likelihood evaluations ($16\,810\,288$) to the number of simulations ($80\,000$). The realistic speed-up with SBI compared to MultiNest is likely closer to this number. 

\begin{deluxetable}{lcc}
\tablecaption{Comparison of the computational cost of MultiNest and SBI. The baseline MultiNest runs use exploratory sampler settings ($10^{3}$ live points, sampling efficiency $0.3$) with importance nested sampling enabled. The upper block compares the number of forward model calls: likelihood evaluations for MultiNest versus number of simulations for SBI. The lower block compares the total inference time: CPU time for MultiNest (dual AMD EPYC 7H12) against CPU (Intel Core i9-13900H) plus GPU (NVIDIA RTX 4060 GPU) time for SBI. \label{tab:cost}}
\tablewidth{\columnwidth}
\tablehead{
  \colhead{} &
  \colhead{\texttt{ST} model} &
  \colhead{\texttt{ST-U} model}
}
\startdata
Likelihood evaluations (MultiNest)                      & $2\,612\,938$ & $30\,226\,982$ \\
Simulations (SBI)                                       & $80\,000$     & $125\,000$    \\
Reduction in forward model calls [\%]                   & $96.9\%$      & $99.6\%$      \\
Speed-up factor                                         & $32.7\times$  & $241.8\times$  \\
\hline
MultiNest CPU time [core hours]                         & $296.4$       & $6207.2$ \\
SBI CPU time, simulation [core hours]                   & $3.3$         & $10.1$         \\
SBI GPU time, training [hours]                          & $2.1$         & $4.0$         \\
Speed-up factor                                         & $54.9\times$  & $440.2\times$  \\
\enddata
\end{deluxetable}

\section{Discussion and conclusions}\label{sec:discussion}
We have developed and tested an SBI framework, based on TSNPE, for neutron star pulse profile parameter inference. Using synthetic NICER data for the \texttt{ST} and \texttt{ST-U} models, we compared this approach to nested sampling in terms of its ability to recover parameters, estimate the evidence, and its computational cost, weighing the trade-offs. Before summarizing our findings, we discuss what the results imply for the method and where its current limits lie.

\subsection{Limitations}\label{subsec:limitations}
The direct application of our current SBI methodology to real observations is limited in three aspects. First, the neutron star models on which the neural density estimator is trained assume a fully ionized hydrogen atmosphere. We did not test alternative compositions, such as the helium atmosphere preferred for PSR~J2124--3358 \citep{Gonzalez-Caniulef2026-J2124}. However, we expect this to only have a minor effect on the current setup, since different atmospheres alter the pulse profiles only slightly.

Second, we assume a phase-independent (unpulsed) background with a power-law photon number spectrum. As discussed in Section~\ref{subsec:xpsi}, the impact of the background depends on the instrument. For future telescopes such as eXTP and NewAthena the predicted background is very low -- $\sim1\times10^{-2}$ c s$^{-1}$ for the eXTP Spectroscopy Focusing Array \citep{Zhang2025}, and $\sim1\times10^{-3}$ c s$^{-1}$ for the NewAthena Wide Field Imager\footnote{The NewAthena Wide Field Imager background files are available here: \url{https://www.mpe.mpg.de/ATHENA-WFI/response_matrices.html}.} -- so we expect its impact to be minor. For NICER, the background is high ($\sim$1 c s$^{-1}$ at best), and is highly variable \citep[e.g.,][]{Remillard2022-3C50}. In this case the background forms a large, uncertain fraction of the counts that a single power law likely cannot capture without biasing the inferred mass and radius, and since our power-law model is also more restrictive than the treatments used in real NICER analyses (Section~\ref{subsec:models}), this regime probably requires further work that remains to be tested. In future work, the background could be treated as a set of nuisance parameters, by letting the expected background counts vary freely in each energy channel. The neural density estimator would then learn to marginalize over them, at the cost of a higher-dimensional inference problem.

The third limitation is that we have considered only single- and two-temperature circular hot regions -- the \texttt{ST} and \texttt{ST-U} models, respectively -- and more complex geometries still need to be tested. More generally, it would be valuable to study how the computational cost of SBI scales with the number of free parameters in the context of PPM.

These three limitations all concern the simulator rather than the methodology, and none of them invalidate the results presented here. 

A further limitation, inherent to TSNPE, is the loss of amortization. The trained network applies to one observation rather than to any observation drawn from the prior. For PPM this matters little, since only a handful of observations are suitable for this technique, but it does mean that for each new observation the network needs to be retrained. As a result, extensive simulation-based calibration or robustness checks of the underlying physics model become more expensive than in an amortized scenario. 

\subsection{Future work}
There are several avenues that would be interesting to explore. To start with, the hot regions of rotation-powered millisecond pulsars are generally modeled as circular or oval spots with uniform temperature, a simplification made to reduce the computational cost of nested sampling. The speed-up offered by SBI opens up the possibility of modeling more physically motivated, complex surface patterns that arise from magnetospheric current models \citep{Gralla17, Lockhart2019, Huang2026}. A second avenue is to apply the SBI approach to accreting millisecond pulsars, whose additional disc and atmosphere components make their models more complex \citep{Dorsman2025-AMPs}, and hence more computationally expensive. A third avenue is to incorporate X-ray polarimetry, which helps constrain geometric parameters \citep{Salmi2025}. As this adds both model complexity and an additional data set, it too would benefit from the speed-up offered by SBI. 

In the future, we aim to improve the framework to the point where it can be applied to real data. Ultimately, the SBI framework should be validated on real observations. A good starting point is the set of rotation-powered millisecond pulsars already analyzed with nested sampling against which the SBI results can be cross-checked. Because the computational cost of SBI is significantly lower than that of nested sampling, it is also well suited for stress testing: exploring many different model configurations and priors to probe the robustness of the inference. 

A further motivation for SBI, not explored here, is that it requires only a forward simulator. Models containing stochastic processes that can be simulated but not written down in closed form are impractical for nested sampling but do not pose any difficulty for SBI. One example is the turbulent accretion flow onto accreting millisecond pulsars \citep{Das2025}. We note, however, that the neural importance sampling estimate of the evidence used here for model comparison relies on an explicit likelihood \citep{Dax2023-IS}, and so cannot be applied in this case. 

\subsection{Summary}
For both the \texttt{ST} and \texttt{ST-U} models, TSNPE recovers the injected parameter values and yields posteriors consistent with those from MultiNest. Because we retain access to the likelihood, we can also estimate the Bayesian evidence within the SBI framework using neural importance sampling. The resulting values agree with MultiNest to within $\sim0.2-0.4$ in $\ln\mathcal{Z}$, at a sample efficiency of $3.09\%$ and $0.40\%$ for \texttt{ST} and \texttt{ST-U} models, respectively. The computational cost of TSNPE is significantly lower: accurate posteriors required $8\times10^{4}$ simulations for \texttt{ST} and $1.25\times10^{5}$ for \texttt{ST-U}, against $2.6\times10^{6}$ and $3.0\times10^{7}$ likelihood evaluations for the corresponding baseline MultiNest runs, which translates to factors of $32.7$ and $241.8$ fewer forward model calls. The gain is larger for the more complex \texttt{ST-U} model, so TSNPE’s advantage over MultiNest becomes more prominent as model complexity grows. These numbers are conservative, as the MultiNest baseline uses exploratory sampler settings; measured against converged sampler settings, the speed-up for the \texttt{ST} model increases from $32.7$ to $210.1$. Finally, SBI also recovers the injected mass and radius (within the 95\% credible interval and in most cases within the 68.3\% interval), and remains well calibrated at all three tested background levels (1\%, 10\%, and 20\%), confirming that the method is robust to the background level.

In this work, we applied SBI to infer parameters from synthetic NICER neutron star pulse profiles. Our results lay the groundwork for extending the method to real data, while the speed-up relative to nested sampling opens the door to more complex models, higher quality data sets expected from future telescopes such as eXTP and NewAthena, as well as to systematic stress testing.

\begin{acknowledgments}
We are grateful to Christoph Weniger for suggesting the use of PCA as a source of summary statistics (Section~\ref{subsec:pca}).
M.H. and A.L.W. acknowledge support from the NWO grant ENW-XL OCENW.XL21.XL21.038 \textit{Probing the phase diagram of Quantum Chromodynamics} under the grant ID \url{doi.org/10.61686/RMOHL62194} (PI: Watts). S.G. is supported by grant ANR-25-CE31-7901-01 (DENSER), and acknowledges the support of the CNES. 
The NWO Domain Science subsidized the use of the national computer facilities in this research under the grant ID \url{https://doi.org/10.61686/OJRAH31001}.

\end{acknowledgments}



\facilities{NICER}

\software{Cython \citep{Behnel2011}, GetDist \citep{Lewis2019}, GNU Scientific Library \citep{Galassi2009}, HEASoft \citep{heasoft2014}, Matplotlib \citep{Hunter2007}, MPI for Python \citep{Dalcin2008}, MultiNest \citep{Feroz2009-Multinest}, nestcheck \citep{Higson2018-nestcheck-joss}, NumPy \citep{Walt2011}, PyMultiNest \citep{Buchner2016-PyMultiNest}, Python/C language \citep{Oliphant2007}, PyTorch \citep{Paszke2019-pytorch}, SBI \citep{BoeltsDeistler2025-sbi}, SciPy \citep{Jones2001-scipy}, and X-PSI \citep{xpsi}.}

\appendix
\section{Posterior distributions at 20\% background}
\begin{figure*}[!]
\centering
\includegraphics[width=1.\textwidth]{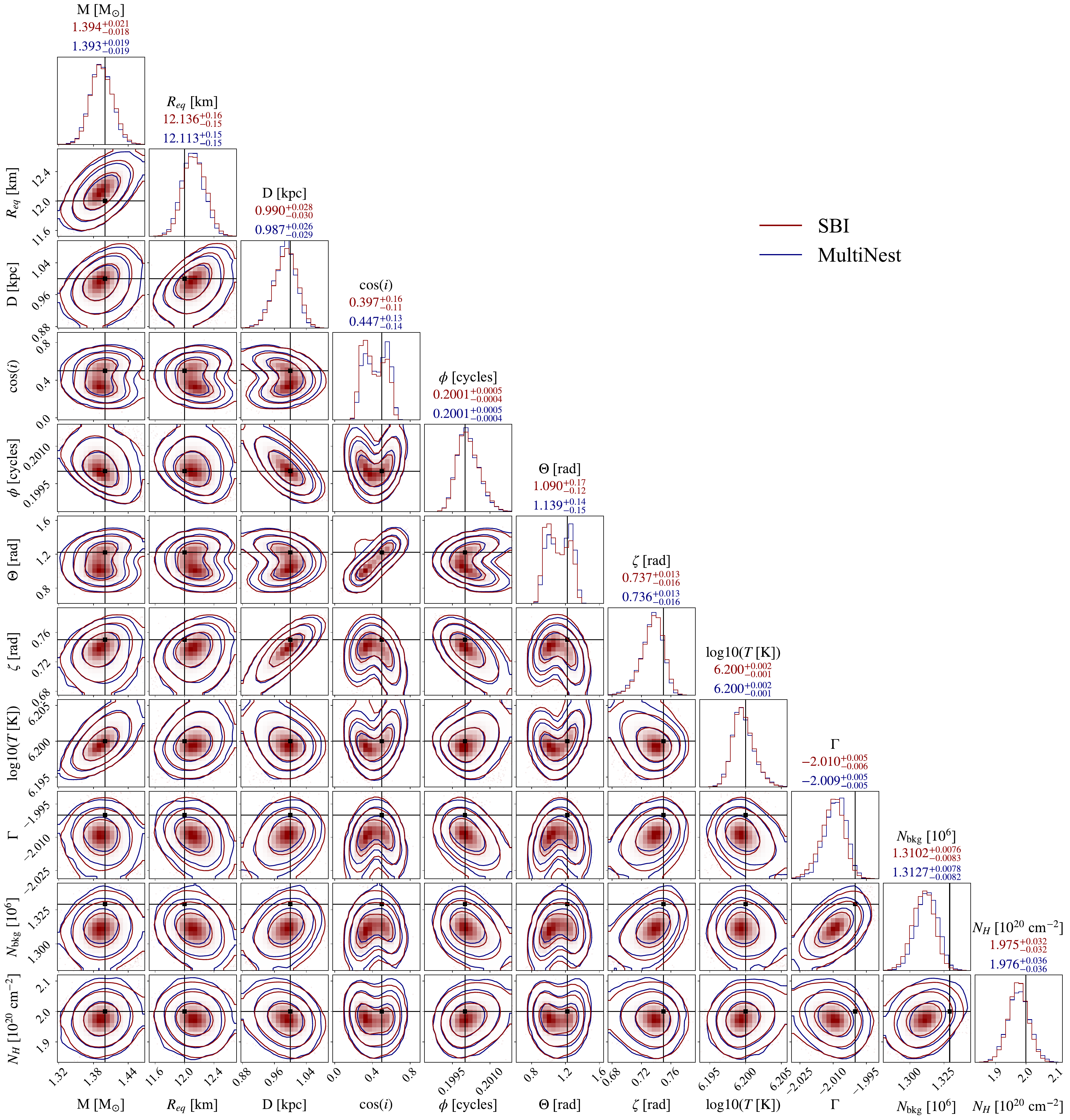}  
\caption{Posterior distributions for the \texttt{ST} model at 20\% background, comparing SBI (dark red) with MultiNest (navy). See Figure~\ref{fig:corner_ST_20bkg_small} for more details about the figure elements.}
\label{fig:corner_ST_20bkg}
\end{figure*}

\begin{figure*}[!]
\centering
\includegraphics[width=1.\textwidth]{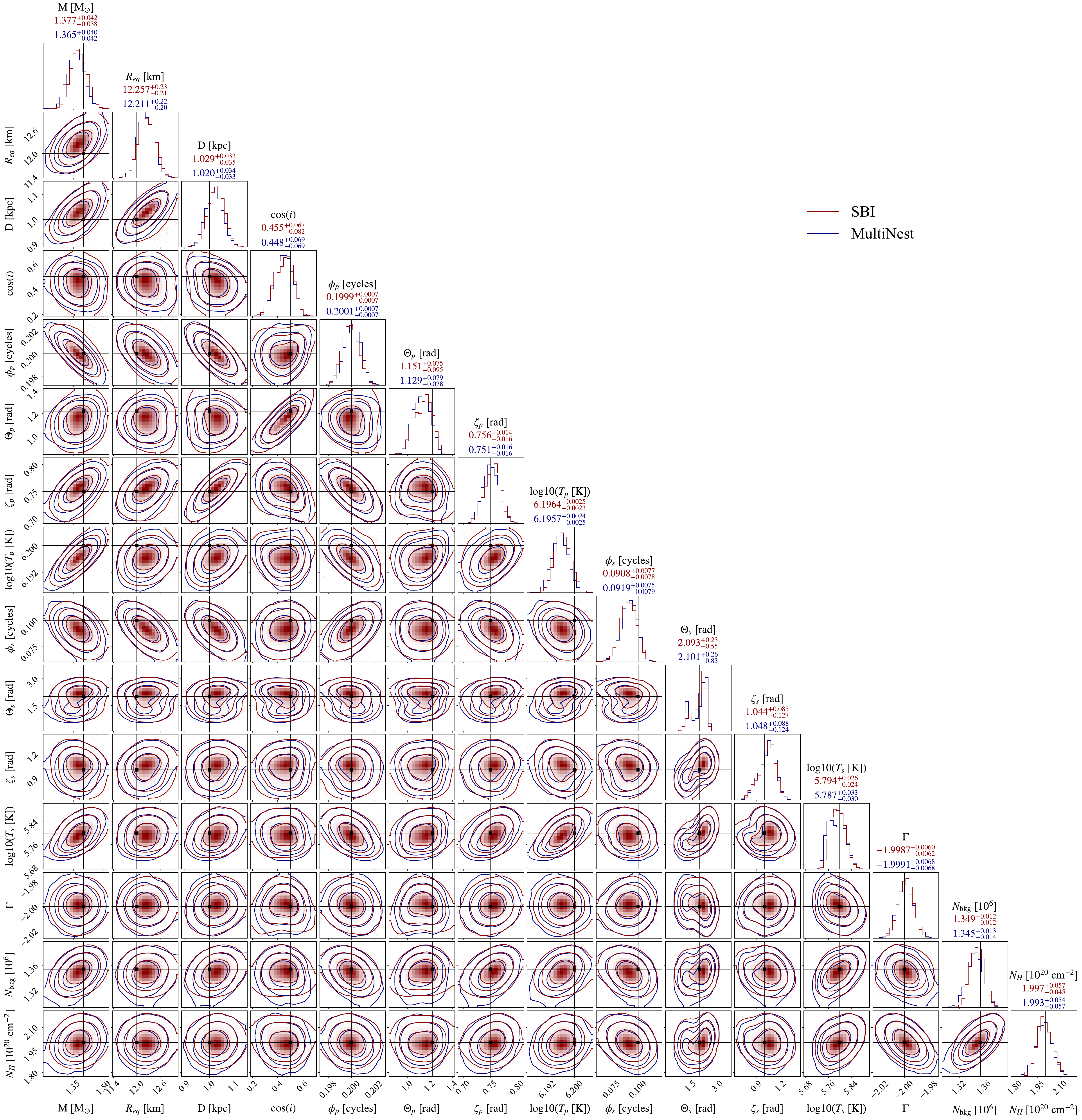}  
\caption{Posterior distributions for the \texttt{ST-U} model at 20\% background, comparing SBI (dark red) with MultiNest (navy). See Figure~\ref{fig:corner_ST_20bkg_small} for more details about the figure elements.}
\label{fig:corner_STU_20bkg}
\end{figure*}

Figures~\ref{fig:corner_ST_20bkg} and~\ref{fig:corner_STU_20bkg} show the posterior distributions for the \texttt{ST} and \texttt{ST-U} models, respectively, obtained from a single observation with 20\% background (see Table~\ref{tab:priors} for the injected values).

\section{Posterior distributions at 1\% background}
\begin{figure*}[!]
\centering
\includegraphics[width=1.\textwidth]{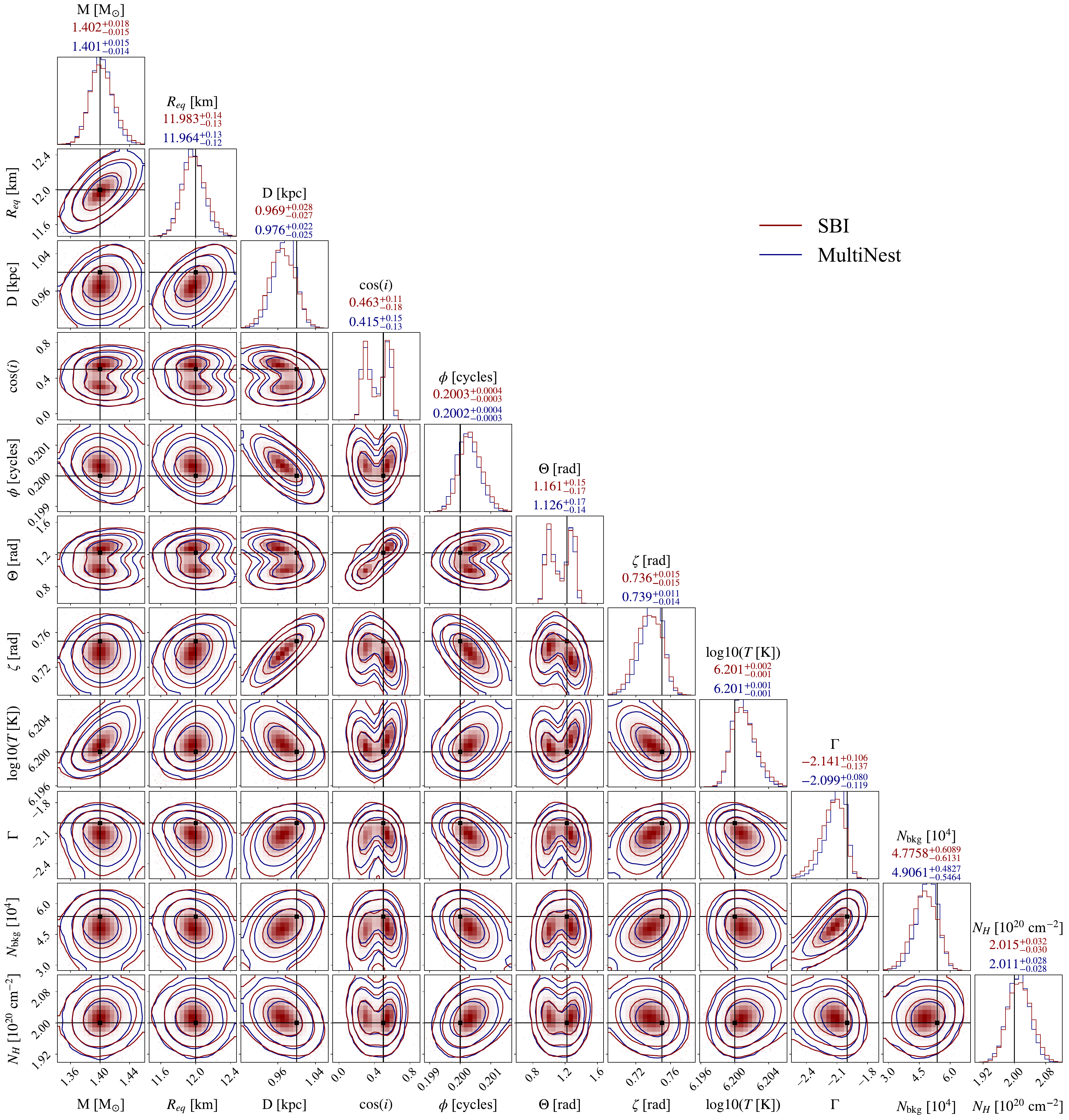}  
\caption{Posterior distributions for the \texttt{ST} model at 1\% background, comparing SBI (dark red) with MultiNest (navy). See Figure~\ref{fig:corner_ST_20bkg_small} for more details about the figure elements.}
\label{fig:corner_ST_1bkg}
\end{figure*}

\begin{figure*}[!]
\centering
\includegraphics[width=1.\textwidth]{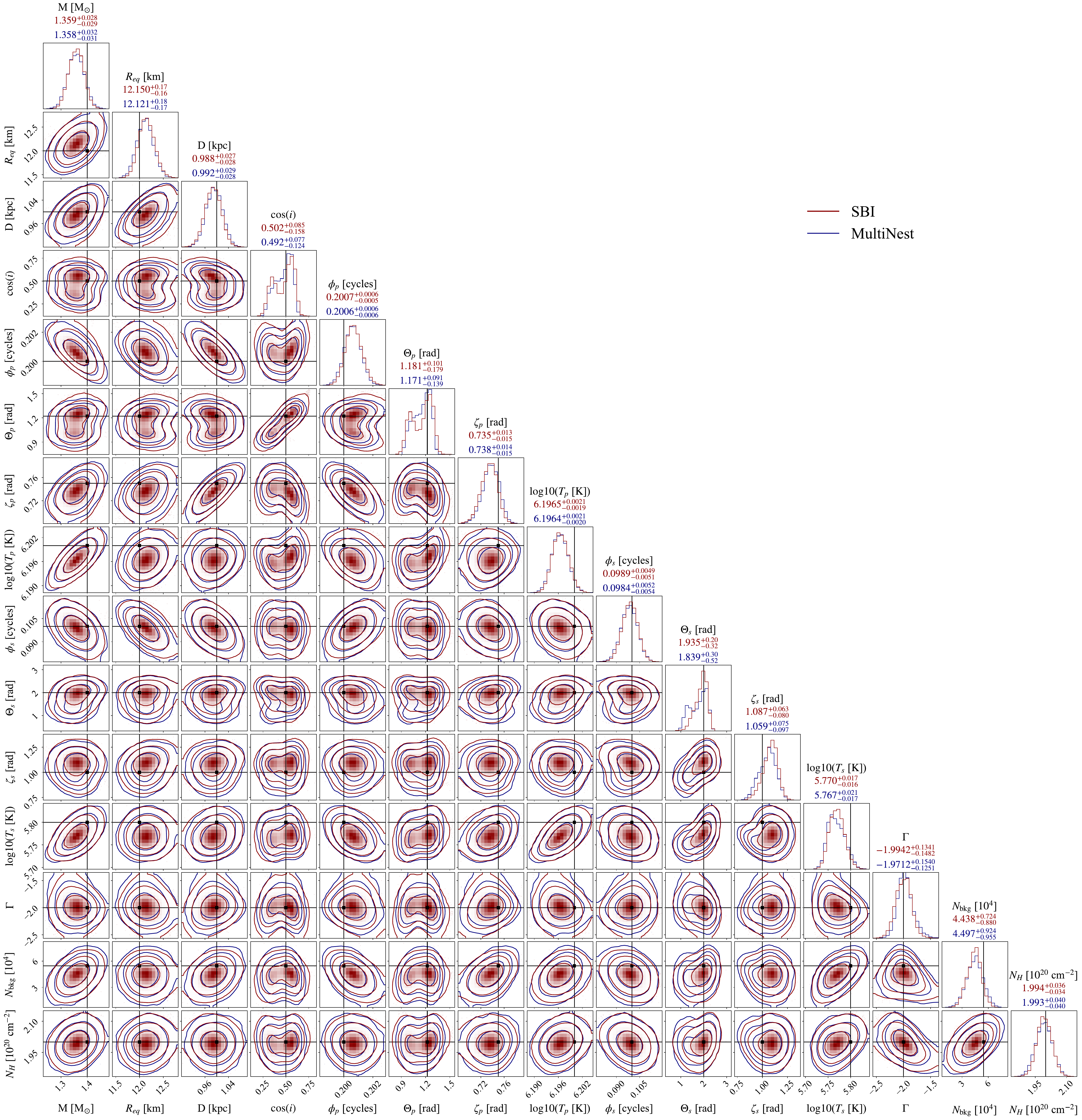}  
\caption{Posterior distributions for the \texttt{ST-U} model at 1\% background, comparing SBI (dark red) with MultiNest (navy). See Figure~\ref{fig:corner_ST_20bkg_small} for more details about the figure elements.}
\label{fig:corner_STU_1bkg}
\end{figure*}

Figures~\ref{fig:corner_ST_1bkg} and~\ref{fig:corner_STU_1bkg} show the posterior distributions for the \texttt{ST} and \texttt{ST-U} models, respectively, obtained from a single observation with 1\% background. The injected values match those listed in Table~\ref{tab:priors}, with the exception of the background counts, which are $N_{\rm bkg}=5.37\times10^{4}$ for \texttt{ST} and $5.47\times10^{4}$ for \texttt{ST-U}.

\section{Evidence convergence}
\begin{figure*}[h!]
\centering
\includegraphics[width=.9\textwidth]{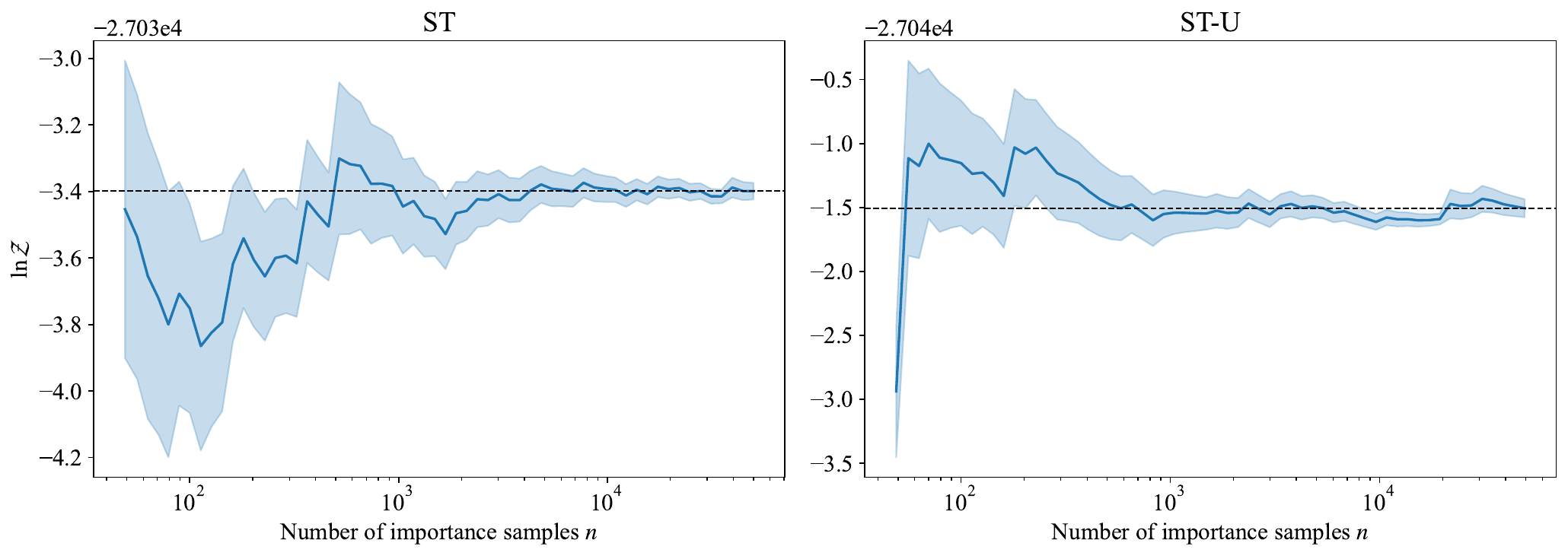}  
\caption{Evidence estimate $\ln\mathcal{Z}$ as a function of the number of importance samples $n$ for the \texttt{ST} (left, $\epsilon = 3.09\%$) and \texttt{ST-U} (right, $\epsilon = 0.40\%$) models at 20\% background.}
\label{fig:evidence_convergence}
\end{figure*}

Figure~\ref{fig:evidence_convergence} shows the evidence estimate $\ln\mathcal{Z}$ as a function of the number of importance samples $n$ for the \texttt{ST} and \texttt{ST-U} models with the SBI method at 20\% background. When the sample efficiency $\epsilon$ is low, the evidence estimate is dominated by a small number of importance samples. An estimate that has not yet converged therefore jumps whenever a high-weight sample is drawn, producing bumps in the evidence estimate (see, for example, Figure 3 of \citealt{Dax2023-IS}). No such bumps appear in Figure~\ref{fig:evidence_convergence}. We conclude that both evidence estimates are converged, despite the somewhat low sample efficiencies of $3.09\%$ for \texttt{ST} and $0.40\%$ for \texttt{ST-U}.

\bibliography{main}{}
\bibliographystyle{aasjournalv7}

\end{document}